\magnification1200
\font\bigbf=cmbx12


\def\pp{\par\hangindent=.125truein \hangafter=1}
\def\aref#1;#2;#3;#4{\pp #1, {\it #2}, {\bf #3}, #4}
\def\abook#1;#2;#3{\pp #1, {\it #2}, #3}
\def\arep#1;#2;#3{\pp #1, #2, #3}
\def\spose#1{\hbox to 0pt{#1\hss}}
\def\simlt{\mathrel{\spose{\lower 3pt\hbox{$\mathchar"218$}}
     \raise 2.0pt\hbox{$\mathchar"13C$}}}
\def\simgt{\mathrel{\spose{\lower 3pt\hbox{$\mathchar"218$}}
     \raise 2.0pt\hbox{$\mathchar"13E$}}}

\def\frac#1/#2{\leavevmode\kern.1em
 \raise.5ex\hbox{\the\scriptfont0 #1}\kern-.1em
 /\kern-.15em\lower.25ex\hbox{\the\scriptfont0 #2}}

\parindent=20pt
\parskip=3pt

\line{\hfil CfPA-95-TH-02}
\line{\hfil January 1995}
\line{\hfil Revised March 1995}
\bigskip

\centerline{\bigbf The COBE Normalization of CMB Anisotropies}
\bigskip

\centerline{Martin White and Emory F. Bunn}
\smallskip

\centerline{\it Center for Particle Astrophysics, University of California,}
\centerline{\it Berkeley, CA 94720--7304}

\noindent
ABSTRACT

\noindent
With the advent of the COBE detection of fluctuations in the Cosmic
Microwave Background radiation, the study of inhomogeneous cosmology
has entered a new phase.  It is now possible to accurately normalize
fluctuations on the largest observable scales, in the linear regime.
In this paper we present a model-independent method of normalizing theories
to the full COBE data.  This technique allows an extremely wide range
of theories to be accurately normalized to COBE in a very simple and
fast way.
We give the best fitting normalization and relative peak likelihoods for
a range of spectral shapes, and discuss the normalization for several popular
theories.
Additionally we present both Bayesian and frequentist measures of the
goodness of fit of a representative range of theories to the COBE data.

\medskip

{\it Subject headings:} cosmic microwave background -- methods: statistical


\bigskip\noindent
{\bf Introduction}
\medskip

Classically, it has been standard practice to normalize models of large-scale
structure at a scale of $\simeq10\,h^{-1}$Mpc, using a quantity related to
the clustering of galaxies
(here the Hubble constant $H_0=100\,h\,{\rm km}{\rm s}^{-1}\,{\rm Mpc}^{-1}$)
measured at the current epoch.
Due to processing of the primordial spectrum and the large amplitude of the
mass fluctuations which galaxies represent, this method of normalization
requires assumptions about the history of the equation of state for matter
inside the horizon, the non-linear evolution of the density field and the
processes of galaxy formation.
One of the key uncertainties is the relationship between the observed
structure and the underlying mass distribution in the universe.
With the COBE DMR detection of Cosmic Microwave Background (CMB)
anisotropies (Smoot et al.~1992), it has become possible
to directly normalize the potential fluctuations at near-horizon scales,
circumventing the problems with the `conventional' normalization.

In an earlier {\it Letter} we presented the normalization of the standard
cold dark matter (CDM) model and a small range of variants
(Bunn, Scott \& White~1995).
In this paper we extend this to a larger class of models, and present a means
for normalizing a whole class of models to the COBE data in a computationally
simple manner.
Throughout we will use the 2-year COBE data 
(Bennett et al.~1994, G{\'o}rski et al.~1994, Wright et al.~1994,
Banday et al.~1994, G{\'o}rski et al.~1995a)
as released by NASA--GSFC.
The normalization in this data
differs from that in the data used by the COBE group by $\sim1\mu$K 
(K.M.~G\'orski, private communication; G\'orski et al., 1995b).

As discussed in Bunn et al.~(1995) and Banday et al.~(1994) there is more
information in the COBE data than just the RMS power measured.
In other words, the COBE data cannot be reduced to a single number without
a significant loss of information.
One way to see this is to notice that there are $\sim90$ eigenvalues of
the signal-to-noise ratio with eigenvalue larger than 1.
An alternative method is to note that the COBE data constrains the amplitude
of the fluctuations over a range of scales, albeit a narrow range. 

In Fig.~1 we show the RMS power in a model which is fittetedd to the COBE data
with one free parameter.  The toy model we have chosen is the so called
Sachs-Wolfe spectrum (Sachs \& Wolfe~1967) which assumes that the observed
temperature fluctuations come purely from the redshifts associated with
climbing out of potentials on the last scattering surface
(for a discussion see e.g.~Peebles~1993; White, Scott \& Silk~1994).
Aside from the normalization of the model (which is fixed by COBE) there is
one free parameter: the spectral slope, denoted $n$, with $n=1$ corresponding
to a scale invariant spectrum.
We notice that the total power is not constant, showing that the normalization
to COBE is sensitive to more than the total RMS fluctuations produced.
Furthermore, the COBE data contain information on the ``shape'' of the power
spectrum, which means that some theories are more likely than others, given
the data.
We will now introduce a simple method for using all of the information in the
COBE data to normalize a wide class of models.

\bigskip\noindent
{\bf Model independent analysis}
\medskip

We have demonstrated that the COBE data contain more information than just
the total power, which can be measured by $\langle Q\rangle$,
$\sigma(10^\circ)$, $\sigma(7^\circ)$, band power, etc.
However, it is still useful to have a method for normalizing a given model to
the data, without having to use the full sky maps containing 6144 pixels at 3
frequencies in each of two channels.
We present here a method which allows one to normalize a large class of
theories (those which can be described by a power spectrum over the limited
range of scales probed by COBE) to the data in a simple manner.

To proceed we notice that theories with gaussian fluctuations (or fluctuations
which are gaussian on COBE scales) can be specified entirely in terms of a
power spectrum.
In CMB studies this is usually expressed as the variance of the multipole
moments, as a function of mode number, i.e.~$\ell(\ell+1)C_\ell$ vs $\ell$,
where the temperature on the sky has been expanded in spherical harmonics
$$
{\Delta T\over T}\left( \theta,\phi \right) \equiv
\sum_{\ell m} a_{\ell m} Y_{\ell m}(\theta,\phi)
\eqno(1)
$$
and we define
$$
\langle a_{\ell m}^{*} a_{\ell' m'}\rangle \equiv
C_\ell\ \delta_{\ell \ell'} \delta_{m m'}
\eqno(2)
$$

For most theories the power spectrum is a smooth function.
Following White~(1994), we will write
$$
D(x) = \ell(\ell+1)C_\ell \qquad {\rm with}\ x=\log_{10}\ell\quad .
\eqno(3)
$$
We can perform a Taylor series expansion of $D(x)$ about some fiducial
point, which we shall take to be $x=1$ ($\ell=10$).
Many theories (see below) can be well approximated by quadratic $D(x)$ over
the relevant range for COBE, roughly $\ell=2$ to $30$, and so we present the
normalizations and likelihoods of quadratic $D(x)$.
We choose to parameterize our quadratics by the (normalized) first and second
derivatives at $x=1$: $D_1'$ and $D_1''$ where
$$
D(x) \simeq D_1 \left( 1 + D_1' (x-1) + {D_1''\over 2} (x-1)^2 \right)
\eqno(4)
$$
(note that $D_1'$ is $1/D_1$ times the derivative of $D(x)$ at $x=1$).
The normalization is then given by quoting $D_1$, or $C_{10}$, for each
$(D_1',D_1'')$ pair, and the goodness of fit is quantified by the relative
likelihood of that shape compared to a featureless, $n=1$, Sachs-Wolfe
spectrum.

We compute the maximum-likelihood normalization for a grid of values
of $D_1'$ and $D_1''$ using the method described in Bunn \&
Sugiyama~(1995) and Bunn et al.~(1995).  We combine the six
publicly-available equatorial DMR sky maps pixel by pixel into a single map by
performing a weighted average, with weights given by the inverse
square of the noise level in each pixel.  (We obtain negligibly
different results when the two 31 GHz maps, which are more sensitive
to Galactic contamination, are excluded.  These maps have high noise
levels, and are therefore automatically given low weights in the
average.)  We excise all pixels whose centers have Galactic latitude
$\left|b\right|<20^\circ$ from the map, leaving 4038 pixels.  

We then estimate the likelihood of getting this particular data set
for each power spectrum.  Rather than computing the likelihood directly,
which would involve repeated inversions of a $4038\times 4038$ matrix,
we first perform a Karhunen-Lo\`eve transform (Karhunen 1947, Thierren~1992)
to ``compress'' the data to a more manageable size.  Specifically,
we choose a set of basis functions $f_i$ defined on the portion of the
sphere outside of the Galactic cut.  (The manner in which these functions
are chosen is described below.)  We then compute the inner product
of the data vector with each of these functions:
$$ 
x_i=\sum_j f_i(\hat{\bf r}_j)\Delta T(\hat{\bf r}_j), \eqno(5) 
$$
where $\hat{\bf r}_j$ is the position of the $j$th pixel, $\Delta T(
\hat{\bf r}_j)$ is the corresponding temperature fluctuation in the
data, and the sum runs over all pixels.  If we assume that the temperature
fluctuations are gaussian, then the projections $x_i$ will be gaussian
as well.  We can therefore compute their likelihood in the usual way:
$$
L\propto \left(\det V\right)^{-1/2}
  \exp\left(-\hbox{$1\over2$} x_i V^{-1}_{ij} x_j\right),
\eqno(6)
$$
where
$V_{ij}\equiv\left\langle x_i x_j\right\rangle$ is the covariance matrix,
which contains contributions from the cosmic signal and the noise:
$$
V=V_{\rm sig}+V_{\rm noise}=(FY)(BCB)(FY)^T+FNF^T,\eqno(7)
$$
where $F_{ij}=f_i(\hat{\bf r}_j)$, $Y_{i\mu}=Y_{\ell m}(\hat{\bf r}_i)$,
$C_{\mu\nu}=C_\ell\delta_{\mu\nu}$, $B$ is the beam pattern,
and $N_{ij}\approx \sigma_i^2\delta_{ij}$ 
is the covariance matrix of the noise in the sky map.
(Here Greek indices stand for pairs of indices $(\ell m)$, with
$\mu=\ell^2+\ell+m+1$.)

The Karhunen-Lo\`eve transform is a prescription for choosing the
basis functions $f_i$, or equivalently the elements of the matrix $F$.
We choose the functions so that the likelihood in eq.~(6)
will have maximal rejection power for incorrect models.  We therefore
want the likelihood $L$ to be, on average, as sharply peaked as
possible about its maximum, or in other words, we want to choose $F_{ij}$
to maximize $\left\langle-L''\right\rangle$, where the primes denote
derivatives with respect to some parameter in our theoretical model and the
derivatives are evaluated at the maximum of $L$.  This optimization
problem reduces to a generalized eigenvalue problem: each row $f_i$ of $F$
satisfies the equation
$$
V_{\rm sig}\vec f=\lambda V_{\rm noise}\vec f.
\eqno(8)
$$
Furthermore, the rows should be chosen to have the maximum eigenvalues 
$\lambda$.  Rows with small values of $\lambda$ probe mostly the distribution
of the noise, with little sensitivity to the cosmological signal.  They
can therefore be omitted from the likelihood estimates with little
loss of information.  We have found that it is necessary to keep only 
the 400 most significant modes. 

Since we have no knowledge of the true monopole and dipole in the sky
map, we marginalize over these modes.
The peak value, width and location of this final marginalized likelihood, as a
function of $D_1$, $D_1'$ and $D_1''$, are the output of the fitting procedure.

Now to find the normalization of any theory, one calculates the large-angle
multipole moments and finds the quadratic which best describes their shape.
Over the range $-0.5\le D_1' \le 0.5$ and $-0.5\le D_1'' \le 3.5$,
the best-fitting amplitude and likelihood are given by the following analytic
forms:
$$
10^{11}C_{10} = 0.8073 + 0.0395 D_1' - 0.0193 D_1''
\eqno(9a)
$$
$$
\ln L = 0.00697 + 1.523 D_1'-0.403 D_1'^2 -0.490 D_1'' -0.0391 D_1'D_1''
+0.00855 D_1''^2
\eqno(9b)
$$
The fitting formula for $C_{10}$ has a worst-case error of $2\%$ and an
average error of $0.4\%$ over this range; the corresponding numbers for
$L$ (not $\ln L$) are $7\%$ and $1.7\%$.
The uncertainty in $C_{10}$ is approximately $15\%$ for all models.  Note
that the COBE data prefer models with positive $D_1'$ and negative
$D_1''$.  The likelihood reaches its maximum at the point
$(D_1',D_1'')=(0.0, -3.0)$, which is beyond the range covered by the
fitting formula.  The peak likelihood for this model is $3.7$ times
the likelihood of a flat Harrison-Zel'dovich model.  Fig.~2 shows
$L$ as a function of $D_1'$ and $D_1''$.

In Tables~1 and 2 we show the best fitting shape parameters for some flat,
low-$\Omega_0$ variants of the CDM model.
The fit of a quadratic to these theories gives an error at the worst fit
multipole (in the range $\ell=2$ to $30$) of about $5\%$, with a typical
error of $\simlt2\%$, showing that such theories are well fit by quadratics
over the range of scales probed by COBE.  To quantify the error introduced
by approximating the power spectrum as a quadratic, we computed likelihood
curves for the {\it worst}-fitting model in our sample using both the true
power spectrum and the quadratic approximation.  The two curves differ
by 11\% in peak likelihood and by 0.5\% in normalization.

Critical ($\Omega_0=1$) CDM models with late reionization are also well
fit by quadratic $D(x)$.
The most plausible, though not the only, ionization history in hierarchical
models of structure formation is standard recombination, followed by full
ionization from some redshift
$z_{*}$ until the present.  The fully ionized phase is due
(perhaps) to radiation from massive stars on scales which go non-linear early.
(See Liddle \& Lyth~1995 for further discussion.)
We find that, over the range of $\ell$ probed by COBE, models which have
$z_{*}\simlt100$ are almost indistinguishable from models with no
reionization, assuming standard big-bang nucleosynthesis values for $\Omega_B$.
There is of course damping on degree scales ($\ell\sim100$), but little change
in the spectrum at smaller $\ell$.  Further, the relative normalization of the
matter and radiation power spectra is the same as in models with standard
recombination.  For a CDM model with $h=0.5$ and $n=1$, the quadratic
parameters are well fit by the formulae
$$\eqalign{
D_1' &= 0.738-0.0307 z_{*}+3.32\times 10^{-4} z_{*}^2-1.06\times 10^{-6}
z_{*}^3\cr
D_1''&= 1.554-0.0483z_{*}+3.67\times 10^{-4}z_{*}^2-7.60\times 10^{-7}
z_{*}^3\cr}
\eqno(10)
$$
over the range $30\le z_{*}\le 110$.

The case of open CDM models is more complicated, since the $C_\ell$ exhibit
features at several scales.
To add to the difficulty, there appear to be several different primordial
spectra one can consider in open universe models.
Some models based on inflationary phases
(Lyth \& Stewart~1990, Ratra \& Peebles~1994, Bucher, et al.~1995)
predict power spectra which show an increase in power near the curvature
radius.  All of these calculations make use of basis functions in which there
is exponential damping of power above the curvature radius; however, this
assumption can be relaxed
(Lyth \& Woszczyna~1995, Yamamoto et al.~1995).
For further discussion of these issues see the appendix.
The open models of Ratra \& Peebles~(1994) have already been fitted to the
2-year COBE data (G{\'o}rski et al.~1995a) and we will not duplicate the
results here.
We mention, however, that the $C_\ell$ for such models can be fitted by cubics
in $\log_{10}\ell$ to the same accuracy that the $\Lambda$CDM models can be
fitted by quadratics.
This increases the dimension of the parameter space and makes tabulating the
results more difficult.  We defer consideration of cubic fits until the
situation with regard to open models is more settled.

Once the 4-year COBE data becomes available, we hope a fitting formula
similar to Eq.~9 (but which goes to sufficiently high order to encompass
almost all theories), could be produced for the benefit of the astrophysics
community.
Such a fit, coded into a subroutine, would allow any theory to be quickly
and accurately fitted to the COBE data.
At present our simple quadratic fit is sufficient for a wide range of theories
of current interest.

\bigskip\noindent
{\bf The goodness of fit}
\medskip

Statistical methods for using a data set like COBE to place constraints on
models generally come in two varieties, Bayesian and frequentist.
Most CMB work, including analyses of the COBE data as well as other
experiments, has taken a Bayesian point of view.
In the Bayesian approach, the probability of observing the actual data is
computed for each model.  This may be denoted $p(D | M)$, meaning
``the probability of the data given the model''.
One then assumes a ``prior'' probability distribution $p(M)$ on the models
and applies Bayes's theorem to produce a ``posterior'' probability distribution
giving the likelihood $p(M|D)$ of the various models given the data:
$$
p(M|D) \propto p(D|M)p(M).
\eqno(11)
$$
The posterior probability distribution tells us how likely each member of our
family of models is compared with any other member, which is what we would
like to know.

The Bayesian approach is perfectly adequate for assessing the {\it relative}
merits of the various models under consideration: with this approach we can
say, for example, that model A is 10 times more likely than model B.
These models may only differ by a normalization or could be drawn from
different cosmologies or structure formation scenarios.
In some cases however we would like to assign {\it absolute} consistency
probabilities to models.
The Bayesian approach is not well suited to answering this sort of question,
and the problem of assigning an absolute consistency probability to a model
is best attacked with frequentist methods.

In the frequentist approach, we choose some goodness-of-fit parameter $\eta$,
and compute its probability distribution over a hypothetical ensemble of
realizations of the model.  We then compute the value of $\eta$ corresponding
to the real data, and determine the probability of finding a value of $\eta$ as
extreme as the observed value in a random member of our ensemble.
We take this probability to be a measure of the consistency of the data with
the model: if the data does not occur often in realizations of the model,
we say the model is ``unlikely'' given the data.

Two points about this technique deserve emphasis.  First, the consistency
probabilities derived in this manner are conceptually quite distinct from
Bayesian likelihoods.  Bayesians and frequentists ask different questions of
their data, and will therefore sometimes get different answers.
We do expect that models which have low Bayesian likelihoods will in general
have poor frequentist consistency probabilities; however, there is no generally
applicable quantitative relation between the two.
Second, it is clear that the success of the frequentist approach depends on
choosing an appropriate goodness-of-fit parameter $\eta$.  For some classes of
problems a standard choice is available; for the problem we consider below,
we are unaware of such a standard.
This is because the measurement of CMB fluctuations involves detecting extra
noise on the sky.  Thus the correlation function, or errors on the
temperatures, which are used in the fit depend on the theory being
tested, unlike normally examined cases of model fitting.

We wish to assign frequentist consistency probabilities to various power
spectra.  The first goodness-of-fit parameter one might think of for this
purpose is a simple $\chi^2$,
$$
\chi^2 \equiv \sum_{i=1}^M \left(x_i\over\sigma_i\right)^2,
\eqno(12)
$$
where $x_i$ is the amplitude of the $i$th element in our eigenmode expansion,
$\sigma_i^2$ is the variance predicted for $x_i$ by our model, and $M$ is
the number of modes in the expansion.
(In order to remove all sensitivity to the monopole and dipole, the eigenmodes
$f_i$ should be orthogonalized with respect to these modes before the
$x_i$ are computed.)
This parameter would be a natural choice if we wished to constrain the
normalization of a model; however, our primary interest is in constraining
the {\it shape} of the power spectrum, and this goodness-of-fit parameter is
not well suited for this purpose.
In fact, given {\it any} power spectrum $C_\ell$, we can choose a normalization
that gives a $\chi^2$ that lies exactly at the median of its probability
distribution, since $\sigma_i$ scales with the normalization of the theory.
We would therefore conclude that for some normalization this model is a
perfectly good fit regardless of the shape of the $C_\ell$.

To focus on the power spectrum, let us consider quantities quadratic in the
amplitude of the eigenmodes.  There is a complication due to the presence of
the galaxy in the COBE maps, which breaks the rotational symmetry of the COBE
sky. This makes it difficult to define a rotationally symmetric quantity
(like $C_\ell$) by summing over azimuthal variable $m$.
However, this is only a technical complication and we can still define a
measure of power by binning the squares of the mode amplitudes in bins
that probe particular angular scales.  We expand each eigenmode $f_i$
in spherical harmonics and compute an ``effective $\ell$'' probed by
that mode by performing a weighted average over $\ell$ with weights given
by the squares of the coefficients of the expansion.  (The modes are generally
quite narrow in $\ell$-space, so the results are not sensitive to the
exact method of computing the effective $\ell$.)  We then sort the 
modes in order of increasing effective $\ell$ (decreasing angular scale).
As it happens, the result of this procedure is almost identical to
sorting the modes in order of decreasing signal-to-noise eigenvalue.
We then compute the quantities
$$
z_i=\sum_{j=(i-1)K+1}^{iK} \left({x_j\over\sigma_j}\right)^2
\eqno(13)
$$
for $1\le i\le M/K$.
We should choose the bin size $K$ to be large enough to reduce the
intrinsic width of the distribution of $z_i$ to a reasonable level,
yet small enough that the mode amplitudes in each bin probe similar
angular scales.  We have adopted $K=10$ as a compromise between these
two considerations.

If our model is correct, then each $z_i$ will be approximately $K$.
If the model is incorrect, then some $z_i$ will be too low, and others will
be too high.
For example, if our model has too little large-scale power, then the
variances $x_j/\sigma_j$ will be greater than 1 for small $j$, and the
first few $z_i$ will tend to be larger than $K$.
We can quantify this observation by defining the goodness-of-fit parameter
$$
\eta = \sum_{i=1}^{M/K}(z_i-K)^2.
\eqno(14)
$$

The $z_i$ for a Harrison-Zel'dovich spectrum and our worst-fitting
model (model 4 of Table~3) are shown in Fig.~3.  (In making
Fig.~3 we chose the coarser bin size $K=20$ rather than $K=10$, to
reduce scatter in the points.)
Note that with our definition the $z_i$ only loosely correspond to $C_\ell$
and depend on the theory.  When $z_i$ is larger than its expected
value, one can conclude that the data have more power than the theory
on the corresponding angular scale; however, there is {\it no} direct
proportionality between $z_i$ and the corresponding $C_\ell$.
Each $z_i$ can be regarded as an estimator of the power spectrum
of the signal and noise {\it combined}.  For small $i$, $z_i$ samples
mostly signal, while the noise dominates for large $i$.  The value
of $z_i$ for large $i$ therefore changes very little as the model
parameters are varied, as can be seen in Fig.~3.  

Since the mode amplitudes $x_i$ are in general correlated, it is
not possible to compute analytically the probability distribution
of $\eta$.  We must therefore resort to Monte Carlo simulations.
For each of several models, we created $1{,}000$ random sky maps.  We
added noise to each pixel by choosing independent gaussian random numbers
with zero mean and standard deviations corresponding to the noise
levels in the real data.  We computed the parameter $\eta$ for each map.
We chose to simulate six different models.  The first four were chosen to span
a range of values of the Bayesian likelihood: we simulated
(1) a Harrison-Zel'dovich spectrum;
the models with the (2) highest and (3) lowest Bayesian likelihoods from
our grid of quadratic power spectra;
and (4) a model with an even lower Bayesian likelihood $L=0.01 L_{HZ}$.
In addition, we chose two models from Table~1 which have identical
cosmological parameters ($\Omega_0=0.1$, $h=0.75$, $n=0.85$), except that
(5) one has only scalar perturbations and one (6) includes tensors in the
ratio $C_2^{T}/C_2^{S}=7(1-n)$.

Our simulation procedure fails to mimic the real COBE data in at least
two ways.  First, the assumption that the noise in different pixels
is independent is not strictly true (Lineweaver et al.~1994).
However, the correlations are quite weak and have been shown to have
negligible effects in analyses similar to ours (Tegmark \& Bunn~1995).
Second, we have not attempted to model the removal of systematic
effects from the data.  However, we expect this to have little
effect on the final results, since the removal of systematic effects
primarily affects the low-$\ell$ multipoles, while proper treatment
of the noise is more important for the high values of $\ell$ where
noise dominates.

It is clear from Table~3 that low Bayesian likelihoods tend to
correspond to poor frequentist consistency probabilities, as expected.
Furthermore, those models with likelihoods of order unity are 
reasonable fits to the data.  This is a very reassuring fact: 
it was perfectly possible {\it a priori} that all the models we have
been considering would prove to be intrinsically poor fits to the data.
Comparison with the consistency probabilities for models 5 and 6 and a
look at Fig.~4 allows one to calibrate the sensitivity of the COBE
data to spectral shape information.  We expect this to improve with the
4-year data, especially at higher $\ell$.

\bigskip\noindent
{\bf The matter power spectrum}
\medskip

The best normalization and the goodness of fit of the temperature
fluctuations for a range of models are given by Eq.~9.
Using these results to normalize the matter power spectrum from the CMB can
present some complications.  In the simplest picture, in which large-angle
CMB anisotropies come purely from potential fluctuations on the last
scattering surface, the relative normalization of the CMB and matter
power spectrum today is straightforward 
(White, Scott \& Silk~1994, Bunn et al.~1995).
In the conventional notation where the radiation power spectrum is given
by
$$
C_\ell = C_2
 {\Gamma\left(\ell+{n-1\over 2}\right)\over\Gamma\left(2+{n-1\over 2}\right)}
 {\Gamma\left(2+{5-n\over 2}\right)\over\Gamma\left(\ell+{5-n\over 2}\right)}.
\eqno(15)
$$
the matter power spectrum for an $\Omega_0=n=1$ CDM universe is
$$\eqalign{
  P(k) & = 2\pi^2\eta_0^4\,A\, k\,T_m^2(k)\cr
  &\simeq 2.5\times 10^{16}A\, (k/h\,{\rm Mpc}^{-1})\, T^2(k)
  \quad (h\,{\rm Mpc}^{-1})^3.  \cr}
\eqno(16)
$$
with $A=3C_2/(4\pi)$.  In models such as CDM this relation works quite well,
as long as matter-radiation equality is sufficiently early ($h$ is not too
low).  Even for $h=0.3$ the relation works at the 4\% level and if $h=1$
the error is $\simlt1\%$.  See Bunn et al.~(1995) for further discussion.

For models with $\Omega_0<1$ the normalization is not so straightforward.
Naively one would think that, for fixed CMB fluctuations at $z=1,000$, one
would have smaller matter fluctuations today.  This is because in an open
or a flat model with a cosmological constant ($\Omega_\Lambda=1-\Omega_0$),
density perturbations stop growing once either the universe becomes curvature
or cosmological constant dominated (respectively).
Curvature domination occurs quite early, and the growth of density fluctuations
$\delta\rho/\rho\equiv\delta$ in an open universe is suppressed (relative to an
$\Omega_0=1$ universe) by a factor $\Omega_0^{0.6}$.
In a flat $\Lambda$ model, the cosmological constant dominates only at late
times and so the growth suppression is a weaker function of the matter content:
$\delta\propto\Omega_0^{0.23}$.
This suppression of growth in an $\Omega_0<1$ universe has often been cited
as ``evidence'' that $\Omega_0$ must be large --- otherwise fluctuations could
not have grown enough to form the structures we observe today.

In fact there are several other effects which come into play when normalizing
the matter power spectrum to the COBE data in a low-$\Omega_0$ model.
The first is that, though the growth in such models is suppressed by
$\Omega_0^{p}$ ($p\simeq0.6$ for open and $0.23$ for $\Lambda$ models;
for a more general formula see Carroll et al.~1992), the potential fluctuations
are proportional to $\Omega_0$.
Hence the CMB fluctuations are even more suppressed than are the density
fluctuations!
So for a fixed COBE normalization the matter fluctuations today are
{\it larger} in a low-$\Omega_0$ universe, and the cosmological constant model
clearly has the most enhancement since the fluctuation growth is the least
suppressed.
In terms of the power spectrum, $P(k)$, we expect for fixed COBE normalization
that $P(k)\propto\delta^2\propto\Omega_0^{2(p-1)}$, as has been pointed out by
Efstathiou, Bond \& White~(1992).

This potential suppression is not the only effect which occurs in
low-$\Omega_0$ universes, although it is the largest.  
Due to the fact that the fluctuations stop growing (or in other words the
potentials decay) at some epoch, there is another contribution to the
large--angle CMB anisotropy measured by COBE.
In addition to the redshift experienced while climbing out of potential
wells on the last scattering surface, photons experience a cumulative
energy change due to the decaying potentials as they travel to the observer.
If the potentials are decaying, the blueshift of a photon falling into
a potential well is not entirely canceled by a redshift when it climbs out.
This leads to a net energy change, which accumulates along the photon path.
This is often called the Integrated Sachs-Wolfe (ISW) effect, to distinguish
it from the more commonly considered redshifting which has become known as the
Sachs-Wolfe effect (both effects were considered in the paper of
Sachs \& Wolfe~1967).
This ISW effect will operate most strongly on scales where the change of
the potential is large over a wavelength, so preferentially on large angles
(Kofman \& Starobinsky~1985).

In $\Lambda$ models the ISW effect can change the relative normalization of
the matter and radiation fluctuations at the $25\%$ level for
$\Omega_0\sim0.3$ (see below).
We show in Fig.~5 how these various effects on the inferred matter power
spectrum normalization scale with $\Omega_0$ in a cosmological constant
universe (the simplest case).
We see the total power is slightly changed, for fixed $C_{10}$, because the
shape of the $C_\ell$ depend on $\Omega_0$.
This affects the goodness of fit with the COBE data
(see Bunn \& Sugiyama~1995 and our Eq.~9b).
The ratio of the large-scale matter normalization to $C_{10}$ is changed by
the ISW contribution to $C_{10}$, the change in the potentials and the
growth of fluctuations from $z=1,000$ to the present.  Over the range
$\Omega_0=0.1$ to $0.5$ one finds
for an $n=1$ spectrum with $C_{10}=10^{-11}$
$$
\lim_{k\rightarrow0} {P(k)\over k} = 1.14\times 10^6 \Omega_0^{-1.35}
\quad (h^{-1}{\rm Mpc})^{4}
\eqno(17)
$$
almost independent of $h$.  This can be compared with the scaling presented
above.
Also the epoch of matter-radiation equality is shifted, which changes the
normalization on smaller scales for fixed large-scale $P(k)$.
Putting these effects together we show the RMS fluctuation on a scale
$0.028\,h{\rm Mpc}^{-1}$ (see below) as a function of $\Omega_0$ in Fig.~6.
The sharp downturn at low $\Omega_0$ is due to a combination of the larger
scale of matter-radiation equality, moving the break in the power spectrum
to smaller $k$, and the photon drag on the baryons having an increased effect
on fluctuation growth for large $\Omega_B/\Omega_0$.
For $\Omega_0\simeq0.3$ the shift in matter-radiation equality and the
scaling of Eq.~(17) roughly cancel, making $\Delta^2$ much less sensitive to
$\Omega_0$ than the individual contributions would suggest.

We note here that the shape of the $C_\ell$ for the tensor
(gravitational wave) modes is largely independent of $\Omega_0$
(Turner, White \& Lidsey~1993).
For this reason the radiation power spectrum of a $\Lambda$ model with
some tilt and a component of tensors can exhibit less curvature at 
$\ell\sim10$ than a purely scalar power spectrum (see Fig.~4).
Since in some inflationary models we expect a non-negligible tensor component
(Davis et al.~1992, but see Liddle \& Lyth~1992, Kolb \& Vadas~1993)
we have computed the tensor $C_\ell$ following Crittenden et al.~(1993) and
give results both including and excluding a significant tensor contribution.
Our results update those of Kofman, Gnedin \& Bahcall~(1993) who also
considered tilted, $\Lambda$CDM models with a component of gravity waves.

In open models, where curvature domination occurs early, much of the
large-angle anisotropy comes from the ISW effect (Hu \& Sugiyama~1995) so
the matter-to-radiation normalization is even more complicated.
For open models the dependence of $P(k\rightarrow0)/C_{10}$ on $\Omega_0$ is
not well fit by a power law, since the shape of the $C_\ell$ on all scales
depends on $\Omega_0$.
Fig.~6 shows the normalization on smaller scales for a model with
$P(q)$ given by Eq.~A3, where $q^2=k^2/(-K)-1$ and $K=H_0^2(\Omega_0-1)$.
[The $C_\ell$ in this model will be similar to those in the inflationary 
models of Bucher et al.~(1995) and Yamamoto et al.~(1995).  
For a discussion of $P(q)$ in open models see the appendix.]
Notice that, for fixed $h$ and CMB normalization, the open models predict
a smaller amplitude for the matter fluctuations today than the
$\Lambda$ models.
We can understand this as a consequence of the earlier onset of curvature
domination than $\Lambda$ domination and the consequently stronger suppression
of fluctuation growth in open models. 

In Tables~1 and 2, and in Fig.~6,we show the normalization of the matter
power spectrum for a range of models where the CMB normalization ($C_{10}$)
is held fixed.
We quote both the value of the RMS density fluctuation at 0.028$h$Mpc$^{-1}$
(large-scale) and $\sigma_8$ (small-scale).
For comparison, Peacock \& Dodds~(1994) give
$$
\Delta^2\left(k=0.028h{\rm Mpc}^{-1}\right) \equiv
  {d\sigma_\rho^2\over d\ln k} =
  \left( 0.0087\pm 0.0023 \right)\Omega_0^{-0.3}.
\eqno(18)
$$
There are many determinations of $\sigma_8$; we quote here those from
Peacock \& Dodds~(1994)
$$
\sigma_8=0.75\Omega_0^{-0.15}
\eqno(19)
$$
and cluster abundances (White, Efstathiou \& Frenk~1993)
$$
\sigma_8=0.57\Omega_0^{-0.56}
\eqno(20)
$$
where the scaling with $\Omega_0$ in both cases refers to models with
$\Omega_0+\Omega_\Lambda=1$.
These values are consistent with those inferred from large-scale flows
(Dekel~1994) and direct observations (Loveday et al.~1992).  In Fig.~7 we
compare these observations with the COBE-normalized values of $\sigma_8$ 
for a range of $\Lambda$CDM models.

\bigskip\noindent
{\bf Conclusions}
\medskip

The COBE data forms a unique and valuable resource for the study of
inhomogeneous cosmology.  To fully exploit this hard won information we
need to go beyond methods of normalizing theories of structure formation
which use only gross properties of the data (such as the RMS fluctuation).
In this paper we have presented a model independent method of parameterizing
the COBE data, and discussed the normalization of the radiation and matter
power spectra for a range of theoretically interesting models.
In addition we considered the question of the goodness of fit of some
commonly adopted models to the data from two complementary statistical
standpoints.


\bigskip\noindent
{\bf Acknowledgments}
\medskip

We thank Douglas Scott for useful conversations and Joe Silk for suggesting
we include a tensor contribution to the $\Lambda$ models.
We would also like to thank Ned Wright for comments on the manuscript and
Section~3 in particular.
The COBE data sets were developed by the NASA Goddard Space Flight
Center under the guidance of the COBE Science Working Group and were
provided by the NSSDC.
This work was supported in part by grants from the NSF and DOE.

\vfill\eject
\bigskip\noindent
{\bf Appendix}
\medskip

In this appendix we make some comments about the fluctuation spectrum
in an open universe.  The material is taken from the work of
Lyth \& Stewart~(1990), Ratra \& Peebles~(1994, and references therein),
Bucher et al.~(1995) and Lyth \& Woszczyna~(1995).
These papers are sophisticated and rigorous treatments of the subject,
and consequently are somewhat lengthy and technical in parts.
Here we try to give a flavor of the problem, building on the rigorous results
of those works.  We will proceed in historical order.

We start by recalling a few points about inflationary cosmology.
In an inflationary theory with $\Omega_0=1$, quantum fluctuations during
inflation give rise to density and potential fluctuations today
(see e.g.~Olive~1990, Mukhanov et al.~1992,
Kolb \& Turner~1990, Linde~1990 and references therein).
The amplitude of the fluctuations is set by the Hubble constant, $H_{\rm inf}$,
when the perturbation crosses out of the horizon, with larger scales (today)
crossing the horizon earlier during inflation.
If the potential of the inflaton (and hence $H_{\rm inf}$) does not change
very much during the time fluctuations on the relevant scales are produced
(exponential inflation) one obtains a scale-invariant spectrum of
fluctuations: $\delta^2\propto k$.
In terms of potential fluctuations, which are the perturbations which enter
the underlying metric and are in some sense more fundamental than the density
perturbations, this corresponds to
$\delta\Phi=$constant\ $\propto H_{\rm inf}/m_{\rm Pl}$.
(This spectrum is therefore described as scale invariant.)
In Lyth \& Stewart~(1990) it was shown that, for scales which entered the
horizon ``early'' when curvature could be neglected, $\delta\Phi=$constant
for $\Omega_0<1$ also.
The transformation from this statement about the fluctuations in $\Phi$ per
logarithmic interval to one about $P(k)$ proceeds in two steps.
First, we relate the density perturbations to the potential fluctuations
through Poisson's equation, which in an open universe reads
(Mukhanov et al.~1992)
$$
\left( k^2 - 3K \right) \delta\Phi_k = 4\pi a^2\rho\delta_k
\eqno(A1)
$$
where the curvature scale is $K=H_0^2(\Omega_0-1)$ as before.
Second, we note that in an open universe, the eigenfunctions of the
operator $\nabla^2$ with eigenvalues $k\ge\sqrt{-K}$ form a complete set
(Lifshitz \& Khalatnikov~1963, Harrison~1970, Abbott \& Schaefer~1986).
Thus we can expand all perturbations in term of these eigenfunctions.
It is convenient to introduce the new variable $q^2=k^2/(-K)-1$ which runs
from 0 to $\infty$.
[In order to obtain the most general gaussian random field in an open
Universe, one must in general include the eigenfunctions with
$0\le k\le\sqrt{-K}$; however, for fluctuations generated by inflation only
modes with $k\ge\sqrt{-K}$ are excited (Lyth \& Woszczyna~1995)].
Recall it is $\delta\Phi_k^2 dk/k$ or $P(k)d^3k$ which gives the physical
fluctuations, so we need to compare the power per logarithmic interval in $k$
to the volume element in the two coordinates
$$
4\pi {dk\over k} = {d^3k\over k^3} = {d^3q\over q(q^2+1)}   .
\eqno(A2)
$$
Writing $k^2-3K\propto q^2+4$, and using the Poisson equation to translate
from potential fluctuations (squared, per $\ln k$) to density fluctuations
(squared, per $d^3q$) we have
$$
P(q) \propto {(q^2+4)^2\over q(q^2+1)}
\eqno(A3)
$$
which is the result of Lyth \& Stewart~(1990).
Ratra \& Peebles~(1994) performed a calculation of fluctuations from a linear
potential in which they showed that this result extends to {\it all} $q$,
not just those which entered the horizon while $\Omega\simeq1$.

In a recent paper, Bucher et al.~(1995) consider an explicit model for open
universe inflation.
[A similar calculation has been done by Yamamoto et al.~(1995)].
In this model the inflaton first gets trapped in a false minimum of the
potential for some time.  During this time the universe inflates exponentially.
The quantum fluctuations in the zero point energy are (power-law) suppressed
by the existence of a mass gap (the inflaton has a mass, since the potential
has curvature at the minimum $V(\varphi)=V_0+m^2\varphi^2/2+\cdots$).
The inflaton then tunnels through the barrier in the standard semi-classical
way (c.f.~nuclear decay) and nucleates a bubble of $\Omega_0\simeq0$ universe.
As the potential rolls slowly from its post-tunneling value to the minimum of
the potential more fluctuations are generated.
The upshot of this, after a strenuous calculation, is that the potential
fluctuations on small scales are $\delta\Phi=$constant, as before.
On larger scales, corresponding to earlier times during inflation, the
potential fluctuations are either enhanced or reduced, depending on the value
of $m$ in the potential.  For the value of $m$ considered in
Bucher et al.~(1995)
one finds an enhancement by an extra factor of $q^{-1}$ on very large scales.
It is worthwhile to stress however that the $C_\ell$ are relatively
insensitive to $q\simlt1$ within reasonable variations in $P(q)$.

The issue of fluctuations in open-universe inflation is not settled.
Further theoretical work is required to determine whether inflation makes
a unique prediction for the fluctuation spectrum in an open Universe,
or whether we must rely on experiments to distinguish among the different
possibilities. 
Should the power spectra rise as $q^{-1}$ or $q^{-2}$ on large scales,
as currently predicted, then models with $\Omega_0$ between $0.1$ and
$0.3$ will be disfavored by the COBE data.

\bigskip\noindent
\vfill\eject
\noindent
{\bf Tables}

\phantom{A}
\vfill

\centerline{ \vbox{ \offinterlineskip \halign{
\strut#&\vrule#&
\hfil#\hfil&\vrule#&
\hfil#\hfil&\vrule$\,$\vrule#&
\hfil#\hfil&\vrule#&
\hfil#\hfil&\vrule#&
\hfil#\hfil&\vrule#&
\hfil#\hfil&\vrule$\,$\vrule#&
\hfil#\hfil&\vrule#&
\hfil#\hfil&\vrule#&
\hfil#\hfil&\vrule#&
\hfil#\hfil&\vrule#\cr
\noalign{\hrule}
&&\quad$\Omega_0$\quad&&\quad$n$\quad&&
\qquad$D_1'$\qquad&&\quad$D_1''$\quad&&
\qquad$\Delta^2$\qquad&&\quad$\sigma_8$\quad&&
\qquad$D_1'$\qquad&&\quad$D_1''$\quad&&
\qquad$\Delta^2$\qquad&&\quad$\sigma_8$\quad&\cr
\noalign{\hrule}
&& 0.50&& 0.85&& -0.079&&  0.476&&  0.027&&  1.45
&& -0.186&&  0.910&&  0.015&&  1.08&\cr
&&     && 0.90&&  0.046&&  0.503&&  0.031&&  1.62
&& -0.043&&  0.785&&  0.020&&  1.31&\cr
&&     && 0.95&&  0.172&&  0.557&&  0.036&&  1.82
&&  0.115&&  0.693&&  0.028&&  1.62&\cr
&&     && 1.00&&  0.298&&  0.635&&  0.040&&  2.05
&&  0.298&&  0.635&&  0.040&&  2.05&\cr
&& 0.40&& 0.85&& -0.110&&  0.600&&  0.030&&  1.29
&& -0.202&&  0.955&&  0.016&&  0.95&\cr
&&     && 0.90&&  0.012&&  0.631&&  0.034&&  1.45
&& -0.065&&  0.853&&  0.022&&  1.15&\cr
&&     && 0.95&&  0.135&&  0.685&&  0.039&&  1.62
&&  0.084&&  0.787&&  0.030&&  1.43&\cr
&&     && 1.00&&  0.258&&  0.767&&  0.045&&  1.82
&&  0.258&&  0.767&&  0.045&&  1.82&\cr
&& 0.30&& 0.85&& -0.159&&  0.853&&  0.032&&  1.10
&& -0.225&&  1.058&&  0.016&&  0.79&\cr
&&     && 0.90&& -0.040&&  0.885&&  0.036&&  1.24
&& -0.097&&  0.991&&  0.022&&  0.97&\cr
&&     && 0.95&&  0.080&&  0.944&&  0.041&&  1.39
&&  0.041&&  0.975&&  0.031&&  1.21&\cr
&&     && 1.00&&  0.201&&  1.022&&  0.047&&  1.56
&&  0.201&&  1.022&&  0.047&&  1.56&\cr
&& 0.20&& 0.85&& -0.240&&  1.338&&  0.029&&  0.82
&& -0.258&&  1.237&&  0.014&&  0.56&\cr
&&     && 0.90&& -0.124&&  1.366&&  0.033&&  0.92
&& -0.143&&  1.229&&  0.019&&  0.69&\cr
&&     && 0.95&& -0.008&&  1.415&&  0.038&&  1.03
&& -0.023&&  1.295&&  0.027&&  0.88&\cr
&&     && 1.00&&  0.108&&  1.485&&  0.043&&  1.15
&&  0.108&&  1.485&&  0.043&&  1.15&\cr
&& 0.10&& 0.85&& -0.342&&  2.455&&  0.014&&  0.33
&& -0.275&&  1.561&&  0.006&&  0.21&\cr
&&     && 0.90&& -0.227&&  2.462&&  0.016&&  0.37
&& -0.176&&  1.673&&  0.008&&  0.26&\cr
&&     && 0.95&& -0.113&&  2.485&&  0.019&&  0.41
&& -0.082&&  1.931&&  0.012&&  0.33&\cr
&&     && 1.00&&  0.000&&  2.527&&  0.022&&  0.46
&&  0.000&&  2.527&&  0.022&&  0.46&\cr
\noalign{\hrule} }} }

\noindent
Table~1: The shape of the radiation power spectrum in a $\Lambda$CDM model
with $h=0.75$.  Also shown is the matter power spectrum normalization with
the radiation normalized to $C_{10}=10^{-11}$.
For the tilted models we show the results with (right columns) and without
(left columns) a gravity wave component with $C_2^{T}/C_2^{S}=7(1-n)$.

\vfill\eject

\phantom{A}
\vfill

\centerline{ \vbox{ \offinterlineskip \halign{
\strut#&\vrule#&
\hfil#\hfil&\vrule#&
\hfil#\hfil&\vrule$\,$\vrule#&
\hfil#\hfil&\vrule#&
\hfil#\hfil&\vrule#&
\hfil#\hfil&\vrule#&
\hfil#\hfil&\vrule$\,$\vrule#&
\hfil#\hfil&\vrule#&
\hfil#\hfil&\vrule#&
\hfil#\hfil&\vrule#&
\hfil#\hfil&\vrule#\cr
\noalign{\hrule}
&&\quad$\Omega_0$\quad&&\quad$n$\quad&&
\qquad$D_1'$\qquad&&\quad$D_1''$\quad&&
\qquad$\Delta^2$\qquad&&\quad$\sigma_8$\quad&&
\qquad$D_1'$\qquad&&\quad$D_1''$\quad&&
\qquad$\Delta^2$\qquad&&\quad$\sigma_8$\quad&\cr
\noalign{\hrule}
&& 0.50&& 0.85&& -0.001&&  0.657&&  0.018&&  0.80
&& -0.123&&  1.039&&  0.010&&  0.60&\cr
&&     && 0.90&&  0.130&&  0.698&&  0.020&&  0.89
&&  0.028&&  0.937&&  0.013&&  0.72&\cr
&&     && 0.95&&  0.262&&  0.769&&  0.023&&  1.00
&&  0.196&&  0.877&&  0.018&&  0.89&\cr
&&     && 1.00&&  0.396&&  0.866&&  0.026&&  1.12
&&  0.396&&  0.866&&  0.026&&  1.12&\cr
&& 0.40&& 0.85&& -0.018&&  0.807&&  0.018&&  0.67
&& -0.130&&  1.099&&  0.010&&  0.50&\cr
&&     && 0.90&&  0.111&&  0.851&&  0.021&&  0.75
&&  0.016&&  1.021&&  0.013&&  0.60&\cr
&&     && 0.95&&  0.240&&  0.924&&  0.024&&  0.84
&&  0.178&&  0.992&&  0.019&&  0.74&\cr
&&     && 1.00&&  0.372&&  1.025&&  0.027&&  0.94
&&  0.372&&  1.025&&  0.027&&  0.94&\cr
&& 0.30&& 0.85&& -0.043&&  1.104&&  0.017&&  0.51
&& -0.140&&  1.222&&  0.009&&  0.36&\cr
&&     && 0.90&&  0.084&&  1.154&&  0.020&&  0.56
&& -0.000&&  1.188&&  0.012&&  0.44&\cr
&&     && 0.95&&  0.212&&  1.227&&  0.022&&  0.63
&&  0.155&&  1.212&&  0.017&&  0.55&\cr
&&     && 1.00&&  0.340&&  1.327&&  0.026&&  0.70
&&  0.340&&  1.327&&  0.026&&  0.70&\cr
&& 0.20&& 0.85&& -0.071&&  1.668&&  0.013&&  0.32
&& -0.147&&  1.436&&  0.006&&  0.22&\cr
&&     && 0.90&&  0.055&&  1.719&&  0.015&&  0.35
&& -0.015&&  1.471&&  0.008&&  0.27&\cr
&&     && 0.95&&  0.180&&  1.788&&  0.017&&  0.39
&&  0.131&&  1.594&&  0.012&&  0.34&\cr
&&     && 1.00&&  0.307&&  1.883&&  0.019&&  0.44
&&  0.307&&  1.883&&  0.019&&  0.44&\cr
&& 0.10&& 0.85&& -0.029&&  2.910&&  0.003&&  0.13
&& -0.112&&  1.806&&  0.001&&  0.08&\cr
&&     && 0.90&&  0.096&&  2.946&&  0.004&&  0.14
&&  0.018&&  1.978&&  0.002&&  0.10&\cr
&&     && 0.95&&  0.222&&  2.998&&  0.004&&  0.15
&&  0.163&&  2.326&&  0.003&&  0.13&\cr
&&     && 1.00&&  0.347&&  3.072&&  0.005&&  0.17
&&  0.347&&  3.072&&  0.005&&  0.17&\cr
\noalign{\hrule} }} }
\noindent
Table~2: As in Table~1 but with $h=0.50$.

\vfill\eject

\phantom{A}
\vfill

\centerline{ \vbox{ \offinterlineskip \halign{
\strut#&\vrule#&
\hfil#\hfil&\vrule#&
\hfil#\hfil&\vrule#&
\hfil#\hfil&\vrule#&
\hfil#\hfil&\vrule#&
\hfil#\hfil&\vrule#\cr
\noalign{\hrule}
&&\quad Model\quad &&\qquad$D_1'$\qquad &&\qquad$D_1''$\qquad&&
\qquad$L$\qquad&&\quad Consistency\quad &\cr
&& && && && && Probability&\cr 
\noalign{\hrule}
&& 1 && 0.4    && -1.6  && 3.29  && 78.0\% &\cr
&& 2 && 0      && 0     && 1.00  && 86.3\% &\cr
&& 3 && -0.5   && 3.5   && 0.091 && 93.0\% &\cr
&& 4 && -2     && 0     && 0.014 && 97.2\% &\cr
&& 5 && -0.342 && 2.455 && 0.186 && 92.4\% &\cr
&& 6 && -0.275 && 1.561 && 0.310 && 91.4\% &\cr
\noalign{\hrule} }} }
\noindent
Table 3: Bayesian and frequentist measures of goodness of fit for six
cosmological models (see text).
The shape parameters $D_1'$ and $D_1''$ are defined in the text.
$L$ denotes the Bayesian peak likelihood, normalized so that a pure
Harrison-Zel'dovich Sachs-Wolfe model has $L=1$.
The consistency probability is the percentage of simulated data sets for which
the goodness-of-fit parameter $\eta$ defined in Eq.~14 is less than the
value found for the real data.

\vfill\eject
\bigskip\noindent
\noindent
{\bf Figure Captions}
\medskip\noindent
Fig.~1:~The value of the RMS power, as measured by $\sigma(7^\circ)$ (solid)
and $\sigma(10^\circ)$ (dashed), for the best fitting, tilted Sachs-Wolfe
models, as a function of $n$.
The fact that the RMS fluctuation of the best fitting model depends on $n$
shows that there is more information in the COBE data than just the RMS power.

\medskip\noindent
Fig.~2:~The likelihood, $L$, as a function of the power spectrum
shape parameters $D_1'$ and $D_1''$.  
The contours range
from $L=0.5$ to $L=3$ in steps of $0.5$, where $L=1$ corresponds to
a flat spectrum $D_1'=D_1''=0$.  For values of $(D_1',D_1'')$ for 
which the power spectrum goes negative over
the range $2\le\ell\le30$ (lower right corner) we have set the likelihood
to zero.

\medskip\noindent
Fig.~3:~The data $z_i$ (see Eq.~13) normalized to a Harrison-Zel'dovich
spectrum (triangles) and our worst fitting model, model 4 of Table~3 
(squares).  The $z_i$ shown were computed with a bin size $K=20$.  The
solid and dashed lines show the
expectation value of each $z_i$ and approximate one-sigma deviations from it.  
(The $z_i$ are only approximately $\chi^2$ distributed and only 
approximately uncorrelated, so these estimated deviations are not precise.)
The
effective $\ell$ values probed by the different modes are indicated
at the top of the figure.  These quantities were computed by expanding
each eigenmode $f_j$ in spherical harmonics and computing the centroid
of the distribution in $\ell$-space, as described in the text. 
Note that with our definition the $z_i$ only loosely correspond to $C_\ell$
and depend on the theory.  At higher $i$ the modes are sampling mostly noise
and therefore change very little as the model parameters are changed.

\medskip\noindent
Fig.~4:~The power spectrum of fluctuations for a $\Lambda$CDM model
with and without the tensor contribution.  The models shown have $h=0.75$,
$\Omega_0=0.1$ and $n=0.85$, 0.90, 0.95 (bottom to top).  The solid lines
are scalar only, while the dashed lines have $C_2^{T}/C_2^{S}=7(1-n)$.
Note the tensor+scalar models have less curvature than the scalar only models,
which makes them a better fitted to the COBE data.

\medskip\noindent
Fig.~5:~An example of how the normalization of the matter power spectrum
depends on $\Omega_0$ in a $\Lambda$CDM model (with $n=1$).
All quantities are normalized to their values at $\Omega_0=0.5$.
The solid line is the RMS temperature fluctuation with $C_{10}$ fixed.
The dotted line shows the ratio of the large-scale matter normalization
($\lim_{k\rightarrow0}P(k)/k$) to $C_{10}$.
The dashed lines show the effect of the shift in matter-radiation equality
on the small-scale normalization $\sigma_8$, holding the large-scale
normalization $\lim_{k\rightarrow0}P(k)/k$ fixed.
We show two models: $h=0.75$ (upper) and $h=0.50$ (lower) with
$\Omega_Bh^2=0.0125$.

\vfill\eject

\medskip\noindent
Fig.~6:~The normalization, $\Delta^2(0.028h{\rm Mpc}^{-1})$, as a function of
$\Omega_0$ for open CDM (dashed) and $\Lambda$CDM (solid) models normalized to
$C_{10}=10^{-11}$.
In both cases the upper curves are for $h=0.75$ and the lower curves are
for $h=0.50$, both with $\Omega_Bh^2=0.0125$.

\medskip\noindent
Fig.~7:~The small scale normalization $\sigma_8$ vs $\Omega_0$ in 
$\Lambda$CDM models, with
normalization set by Eq.~9.  The dashed lines assume all the contribution to
the temperature fluctuations measured by COBE come from scalar perturbations;
the dotted lines are scalars + tensors with $C_2^T/C_2^S=7(1-n)$.
Slope ($n$) increases from 0.85 to 1.00 in steps of 0.05 with lowest $n$ being
lowest $\sigma_8$.  The two solid lines are two observational determinations of
$\sigma_8$, the top line from cluster abundances and the bottom line from large
scale structure (i.e.~Eqs.~19, 20).

\vfill\eject
\bigskip
\frenchspacing
\hoffset=0.25truein
\hsize=6.25truein
\parindent=-0.25truein
{\bf References}
\aref Abbott, L. F., \& Schaefer, R. K., 1986;ApJ;308;546
\aref Banday, A.J. et al. 1994;ApJ;436;L99
\aref Bennett, C. L., et al. 1994;ApJ;436;423
\arep Bucher, M., Goldhaber, A., \& Turok, N., 1995;Phys. Rev.;in press
\aref Bunn, E., Scott, D., \& White, M., 1995;ApJ;441;L9
\aref Bunn, E., \& Sugiyama, N., 1995;ApJ;446;49
\aref Carroll, S. M., Press, W. H., \& Turner, E. L., 1992;Ann. Rev.
A. \& Astrophys;30;499
\aref Crittenden, R., Bond, J. R., Davis, R. L., Efstathiou, G., \&
Steinhardt, P. J., 1993;Phys. Rev. Lett.;71;324
\aref Davis, R. L., Hodges, H. M., Smoot, G. F., Steinhardt, P. J.,
Turner, M. S., 1992; Phys. Rev. Lett.;69;1856 (erratum: 70:1733)
\aref Efstathiou, G., Bond, J. R., \& White, S. D. M., 1992;MNRAS;258;$1\,$p
\aref G{\'o}rski, K. M., et al. 1994;ApJ;430;L89
\arep G{\'o}rski, K. M., et al. 1995a;COBE;preprint
\arep G{\'o}rski, K. M., et al. 1995b;in;preparation
\aref Hu, W. \& Sugiyama, N. 1995;ApJ;444;489
\abook Karhunen, K. 1947;\"Uber lineare Methoden in der
Wahrschleinlichkeitsrechnung; Helsinki: Kirjapaino oy.~sana
\aref Kofman, L., Gnedin, N. Y., \& Bahcall, N. A., 1993;ApJ;413;1
\aref Kofman, L., \& Starobinsky, A., 1985;Sov. Astron. Lett.;11;271
\aref Kolb, E. W., \& Vadas, S., 1993;Phys. Rev.;D50;2479
\abook Kolb, E. W., \& Turner, M. S., 1990;The Early Universe;Addison Wesley
\aref Harrison, E. R., 1970;Phys. Rev.;D1;2726
\aref Liddle, A. R., \& Lyth, D. H., 1992;Phys. Lett.;B291;391
\arep Liddle, A. R., \& Lyth, D. H., 1995;MNRAS;in press
\aref Lifshitz, E. M., \& Khalatnikov, I. M., 1963;Adv. Phys.;12;185
\abook Linde, A., 1990;Particle Physics and Inflationary Cosmology;Harwood
Academic
\aref Lineweaver, C.H., Smoot, G.F., Bennett, C.L., Wright, E.L., Tenorio, L.,
Kogut, A., Keegstra, P.B., Hinshaw, G., \& Banday, A.J., 1994;ApJ;436;452
\aref Loveday, J. S., Efstathiou, G., Peterson, B. A. \& Maddox, S. J.,
1992;ApJ;400;L43
\aref Lyth, D. H. \& Stewart, E., 1990;Phys. Lett.;B252;336
\arep Lyth, D. H. \& Woszczyna, A., 1995;Lancaster;preprint
\aref Mukhanov, V. F., Feldman, H. A., \& Brandenberger, R. H.,
1992;Phys. Rep.;215;203
\aref Olive, K., 1990;Phys. Rep.;190;307
\aref Peacock, J. A. \& Dodds, S. J., 1994;MNRAS;267;1020
\abook Peebles, P. J. E., 1993;Principles of Physical Cosmology;Princeton UP
\aref Ratra, B., \& Peebles, P. J. E., 1994;ApJ;432;L5
\aref Sachs, R. K., \& Wolfe, A. M., 1967;Ap. J.;147;73
\aref Smoot, G., et al., 1992;ApJ;396;L1
\arep Tegmark, M. \& Bunn, E.F., 1995;Berkeley;preprint
\abook Thierren, C.W., 1992;Discrete Random Signals and Statistical Signal
Processing;Prentice-Hall
\aref Turner, M. S., White, M., \& Lidsey, J. E., 1993;Phys. Rev.;D48;4613
\aref White, M., 1994;Astron. \& Astrophys.;290;L1
\aref White, M., Scott, D., \& Silk, J., 1994;Ann. Rev.  A.
\& Astrophys.;32;319
\aref White, S.D.M., Efstathiou, G., \& Frenk C. 1993;MNRAS;262;1023
\aref Wright, E. L., Smoot, G. F., Bennett, C. L., \& Lubin, P. M.,
1994;ApJ;436;443
\arep Yamamoto, K., Sasaki, M., \& Tanaka, T., 1995;Kyoto;preprint

\bye